\documentclass[a4paper]{jpconf}
\usepackage{graphicx}
\usepackage{hyperref}
\usepackage{amssymb, amsmath}
\begin{document}
\title{Scaling laws during collapse of a homopolymer: Lattice versus off-lattice}

\author{Suman Majumder, Henrik Christiansen and Wolfhard Janke}

\address{Institut f\"ur Theoretische Physik, Universit\"at Leipzig, 
Postfach 100\,920, 04009 Leipzig, Germany}

\ead{suman.majumder@itp.uni-leipzig.de, henrik.christiansen@itp.uni-leipzig.de,wolfhard.janke@itp.uni-leipzig.de}

\begin{abstract}
We present comparative results from simulations of a lattice and an off-lattice model of a homopolymer, in the context 
of kinetics of the collapse transition. Scaling laws related to the collapse time, cluster coarsening 
and aging behavior are compared. Although in both models the cluster growth is independent 
of temperature, the related exponents turn out to be different. Conversely, the aging and associated scaling 
properties are found to be universal, with the nonequilibrium autocorrelation exponent obeying a recently derived bound.
\end{abstract}
\section{Introduction}
Understanding of the nonequilibrium dynamics of coarsening of particle and spin systems is quite developed \cite{Bray} and still is a 
topic of interest, especially in systems under special constraints \cite{Suman_nano}. Even though the pathways 
of collapse of a homopolymer, upon a transfer from a good to a poor solvent, bears resemblance to coarsening processes, the kinetics 
of the process has rarely been looked upon from that point of view until only recently 
\cite{Suman_EPL,Suman_aging,Suman_JPCM,Suman_SM,Henrik}. The pioneer work on collapse kinetics goes back to 
the ``sausage'' model of de Gennes \cite{deGennes}. However, simulation results could rather be 
explained by the ``pearl-necklace'' picture of Halperin and Goldbart (HG) \cite{Halperin}. Those studies were 
motivated to understand the scaling of the collapse time, $\tau_c$, with the size of the polymer, $N$, 
using the form $\tau_c \sim N^z$. The exponent $z$ naturally depends on the dynamics of the simulations, 
and so far no agreement has been reached regarding its value.
\par
According to HG, collapse occurs in three stages. The initial stage is the formation of 
nascent clusters of monomers. In the next stage, these clusters grow by pulling more monomers from the chain 
until they eventually coalesce with each other giving rise to a single cluster. In the 
final stage, monomers within the cluster reorganize to form a compact globule at equilibrium. Clearly, 
the second stage of the collapse, i.e., the cluster-growth stage can be identified with the usual coarsening processes. 
Consequently, it is found to be a scaling phenomenon where the average cluster size $C_s(t)$ at time $t$ 
follows a power law $C_s(t)\sim t^{\alpha_c}$, with $\alpha_c$ as the growth exponent. 
\par
Following the same analogy, it has been established \cite{Suman_aging} that one observes aging and 
related scaling during the collapse. To probe aging \cite{Zannetti,Henkel} one defines 
a two-time autocorrelation function  
\begin{eqnarray}\label{auto_cor}
 C(t,t_w)=\langle O_i(t)O_i(t_w) \rangle - \langle O_i(t) \rangle \langle O_i(t_w) \rangle,
\end{eqnarray}
where $t$ and $t_w$ ($\ll t$) are the observation and waiting times, respectively, and $O_i$ is a parameter 
that reflects the spatio-temporal changes in the physical process, e.g., the time- and space-dependent order parameter 
for ferromagnetic ordering. Aging is manifested by the slower decay of $C(t,t_w)$ with increasing $t_w$ and the 
corresponding scaling of $C(t,t_w)$ is given as \cite{Zannetti,Henkel}
\begin{eqnarray}\label{FH-bound}
C(t,t_w) =A_Cx_c^{-\lambda_C},
\end{eqnarray}
where $x_c=\ell/\ell_w$ is the ratio of growing length scales at time $t$ and $t_w$, as in
ferromagnetic ordering \cite{Zannetti,Henkel}, and $\lambda_C$ is the nonequilibrium 
autocorrelation exponent. As for the collapse of a polymer 
\cite{Suman_EPL,Suman_SM,Henrik} the relevant length scale is the 
cluster size $C_s(t)\equiv \ell(t)^d$ (where $d$ is the space dimension), $x_c=C_s(t)/C_s(t_w)$ is 
an obvious choice. In analogy with a bound on $\lambda_C$ popular in spin systems \cite{Fisher}, 
for the collapse of a polymer, too, there exists a $d$-dependent bound \cite{Suman_aging}  
\begin{eqnarray}\label{poly-bound}
 (\nu d-1) \le \lambda_C \le 2(\nu d-1),
\end{eqnarray}
where $\nu$ is the critical exponent related to the size of the polymer in the extended state, i.e., 
$R_g \sim N^\nu$, with $R_g$ being the radius of gyration.
\par
In this article, we present a comparative picture of the above scaling laws, viz., scaling of the 
collapse time, the cluster growth and the autocorrelation functions, from results involving an off-lattice model 
(OLM) and a lattice model (LM) via Monte Carlo (MC) simulations. 
\section{Models and Methods} 
For OLM, we opt for the bead-spring model of a flexible homopolymer in $d=3$ where 
bonds between successive monomers are maintained via the standard 
finitely extensible non-linear elastic (FENE) potential 
\begin{eqnarray}\label{FENE}
E_{\rm{FENE}}(r_{ii+1})=-(K/2) R^2 \ln [1-((r_{ii+1}-r_0)/R )^2 ],
\end{eqnarray}
with $K=40$, $r_0=0.7$ and $R=0.3$. The nonbonded interaction energy is modeled by
$E_{\rm {nb}}(r_{ij})=E_{\rm {LJ}}({\rm{min}}(r_{ij},r_c))-E_{\rm {LJ}}(r_c)$, where 
$E_{\rm {LJ}}(r)=4\epsilon [ (\sigma/r )^{12} - ( \sigma/r )^{6} ]$
is the standard Lennard-Jones (LJ) potential with $\sigma =r_0/2^{1/6}$ as the diameter of the monomers, 
$\epsilon(=1)$ as the interaction strength and $r_c$ $=2.5\sigma$ as the cut-off radius.
\par
For LM, we consider a variant of the interactive self-avoiding walk on a simple cubic lattice, where each lattice site can be 
occupied by a single monomer. The Hamiltonian is given by 
\begin{equation}\label{hamiltonian}
H=-\frac{1}{2} \sum_{i \ne j,  j \pm 1} w(r_{ij}),~~ \textrm{where}~~  
w(r_{ij})=\begin{cases} J & r_{ij} = 1 \\ 0 & \text{else}\end{cases}.
\end{equation}
In Eq.\ \eqref{hamiltonian}, $r_{ij}$ is the Euclidean distance between two nonbonded monomers $i$ and $j$, 
$w(r_{ij})$ is an interaction parameter that considers only nearest neighbors, and $J(=1)$ is the interaction strength. 
We allow a fluctuation in the bond length by considering diagonal bonds, i.e., the possible bond 
lengths are $1$, $\sqrt{2}$ and $\sqrt{3}$. The model has its origin in the bond-fluctuation model \cite{Carmesin} and 
has been independently studied \cite{Shaffer,Dotera} for equilibrium properties. 
\par
We introduce the dynamics in the models via Markov chain MC simulations \cite{Landau_book}. For both models we apply 
local moves, i.e., after selecting a monomer randomly, for OLM, we shift it to a position randomly chosen within 
[$-\sigma/10 :\sigma/10$] of its current position, and for LM, to another lattice site such that the bond-connectivity constraint 
and the excluded-volume criterion are preserved. For details on the allowed moves in LM we refer to Refs.\ \cite{Henrik,Shaffer,Dotera}. 
A trial move is accepted or rejected following the Metropolis algorithm with Boltzmann criterion. One Monte Carlo step (MCS) 
consists of $N$ (where $N$ is the number of monomers in the chain) such attempted moves, effectively setting the time scale. 
\par
The collapse transition temperature is $T_{\theta}(N \rightarrow \infty) \approx 2.65 \epsilon/k_B$ \cite{Suman_SM} 
and $\approx 4.0J/k_B$ \cite{Henrik}, respectively, for OLM and LM. The unit of temperature is $\epsilon/k_B$ or $J/k_B$, where 
the Boltzmann factor $k_B$ is set to unity. We prepare initial conformations of the polymers at high temperatures $T_h \approx 1.5T_{\theta}$ 
that mimics an extended coil phase and then quench it to temperatures $T_q < T_{\theta}$. Since LM is computationally less 
expensive than OLM, for LM we simulate polymers up to $N=4096$ whereas for OLM the longest polymer we simulate has $N=724$. 
All the results presented (except the snapshots) are averaged over at least $300$ different initial realizations.
\begin{figure}[t!]
\centering
\includegraphics*[width=0.43\textwidth]{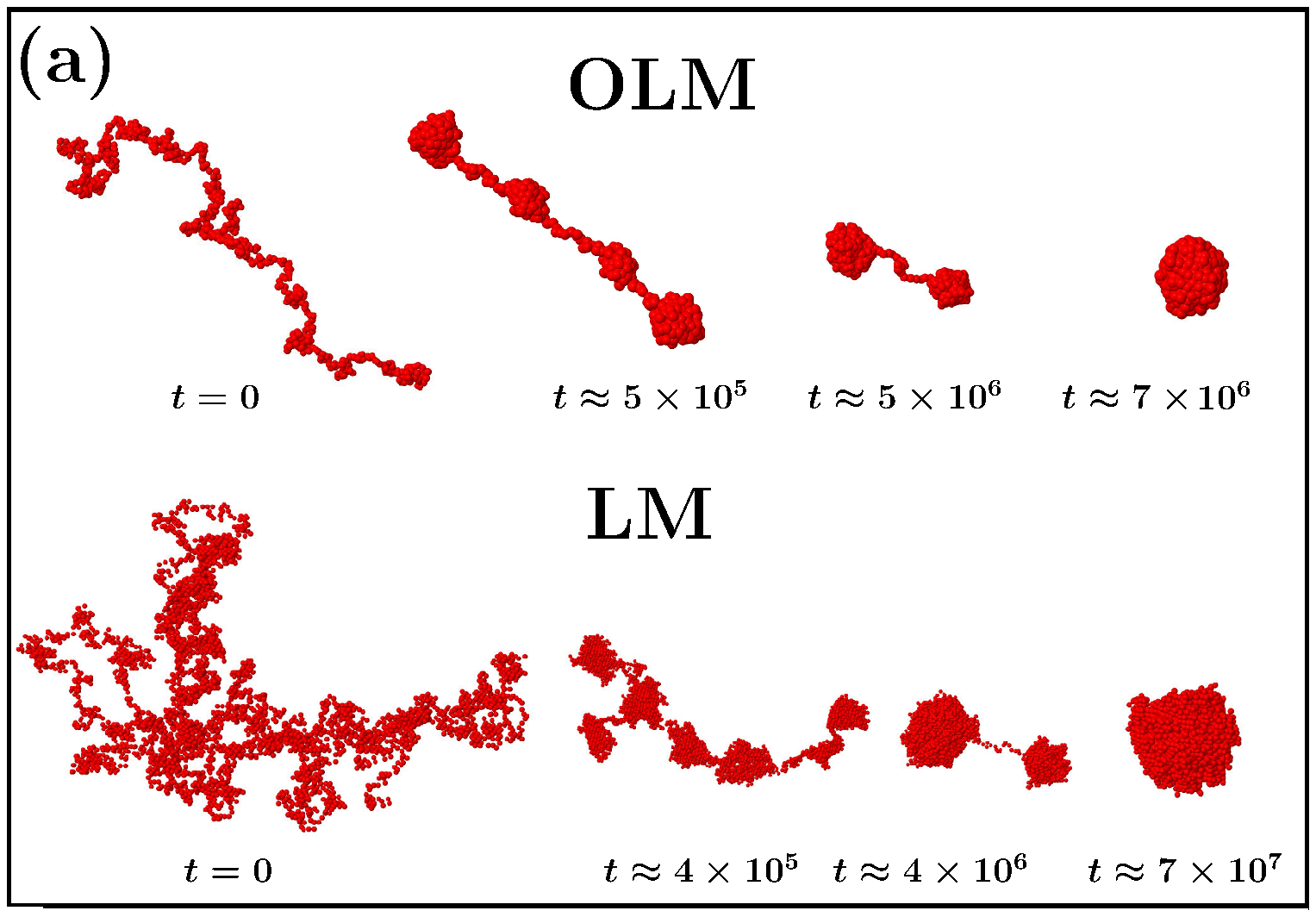}
~~~\includegraphics*[width=0.51\textwidth]{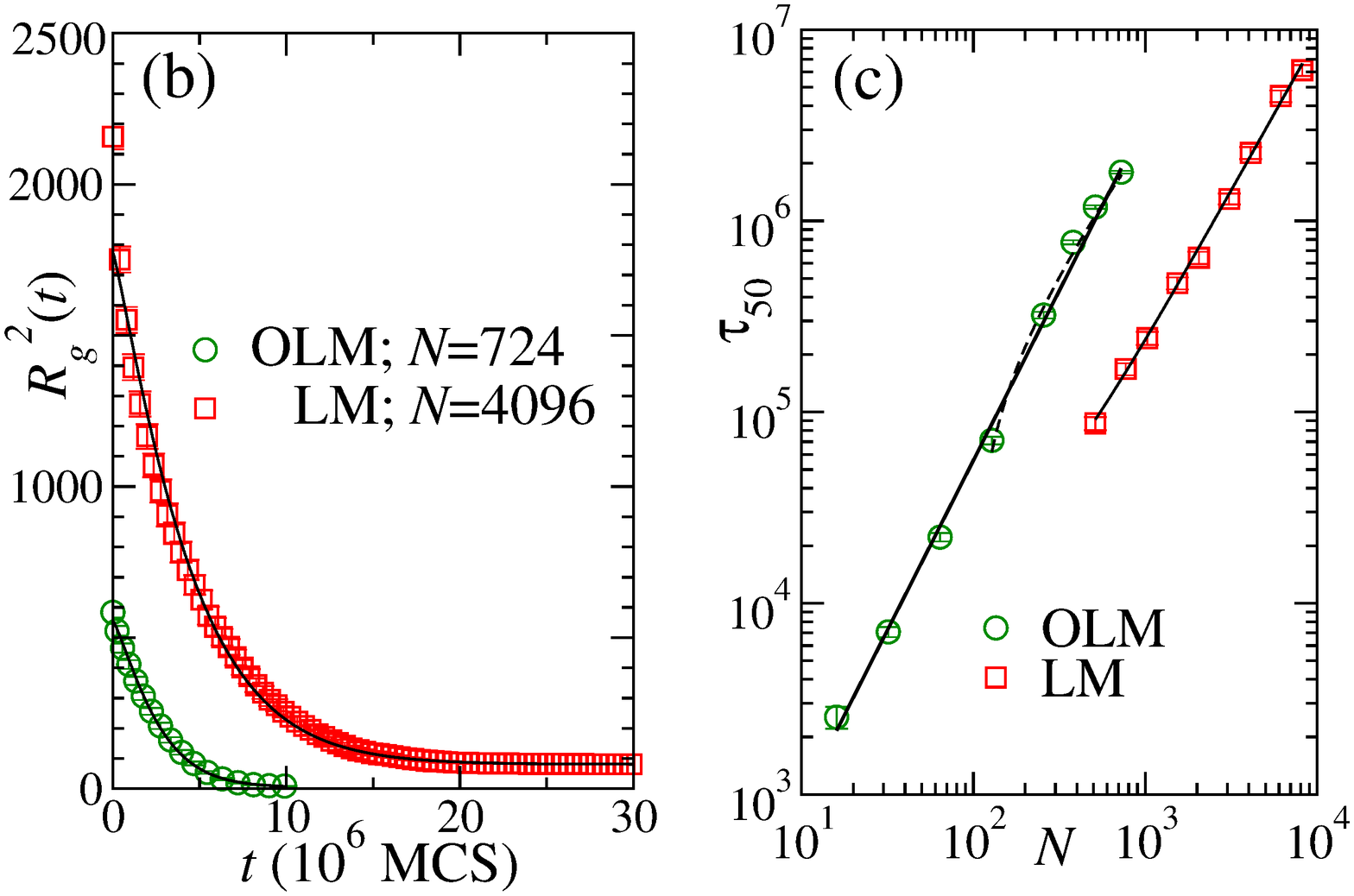}
\caption{\label{fig1}(a) Time evolution snapshots of the collapse of a homopolymer, 
after being quenched from an extended coil phase to a temperature, $T_q=1$ for OLM, 
and $T_q=2.5$ for LM, in the globular phase. (b) Decay of the squared radius of gyration, $R_g^2(t)$, 
with time. The solid lines are fits to a 
stretched exponential form described in the text. (c) Plot of collapse time, $\tau_{50}$, as function of $N$. 
The solid lines are fits to the form \eqref{tau_N}. The dashed line is a fit of the OLM data for $N\ge128$, 
to the form \eqref{tau_N} by fixing $z=1$.}
\end{figure}
\section{Results}\label{Results}
In Fig.\ \ref{fig1} (a) we present snapshots showing the sequence of events during the collapse, 
for $N=724$ and $4096$, respectively, for OLM and LM. Both models provide the same phenomenological picture, 
i.e., initial formation of tiny clusters followed by coarsening of those clusters to form bigger ones and 
eventually a single cluster. This is in agreement with the ``pearl-necklace'' picture of HG. 
However, a careful observation reveals that in LM the coarsening occurs not only along the chain 
but also from lateral branches, which merely is an effect of using a much longer chain for LM. 
\par
As a first step to understand the kinetics, one observes the decay of the squared radius of gyration, 
$R_{g}^2=\sum\limits_{i,j}(r_i-r_j)^2/2N^2$, as shown in Fig.\ \ref{fig1} (b). 
Although this does not provide any detailed information about the stepwise collapse of the polymer, 
one can extract a measure of the collapse time, $\tau_c$, by fitting the decay of $R_g^2(t)$ to a 
stretched exponential function $R_{g}^2=a_0+a_1\exp[-(t/\tau_c)^{\beta}]$, represented by the solid lines 
in Fig.\ \ref{fig1} (b). An elaborate description of such fitting can be found elsewhere 
\cite{Suman_SM,Henrik}. In addition, we also measure the collapse time, $\tau_{50}$, as the time when $R_g^2(t)$ has 
decayed to half of its total decay, i.e., $\left[R_g^2(0)-R_g^2(\infty)\right]/2$. In Fig.\ \ref{fig1} (c), 
we show  the variation of $\tau_{50}$ with $N$. Data for both the models show a power-law scaling, which can 
be quantified using the form 
\begin{eqnarray}\label{tau_N}
\tau_c=B N^z+\tau_0,
\end{eqnarray}
where $B$ is a nontrivial constant that depends on the quench temperature $T_q$, $z$ is the 
corresponding dynamic critical exponent, and the offset $\tau_0$ comes from finite-size corrections. 
For LM a fitting [shown by the solid line in Fig.\ \ref{fig1} (c)] with the form \eqref{tau_N} yields $z=1.61(5)$ 
and is almost insensitive to the chosen range. However, for 
OLM the fitting is sensitive to the chosen range. While using the whole range of data provides $z=1.80(6)$ [shown by the 
solid line in Fig.\ \ref{fig1} (c)], fitting only the data for $N \ge128$ yields $z =1.20(9)$. In this regard, 
a linear fit [$z=1$ in \eqref{tau_N}], shown by the dashed line, also cannot be ruled out \cite{Suman_SM}. 
\begin{figure}[t!]
\centering
\includegraphics*[width=0.93\textwidth]{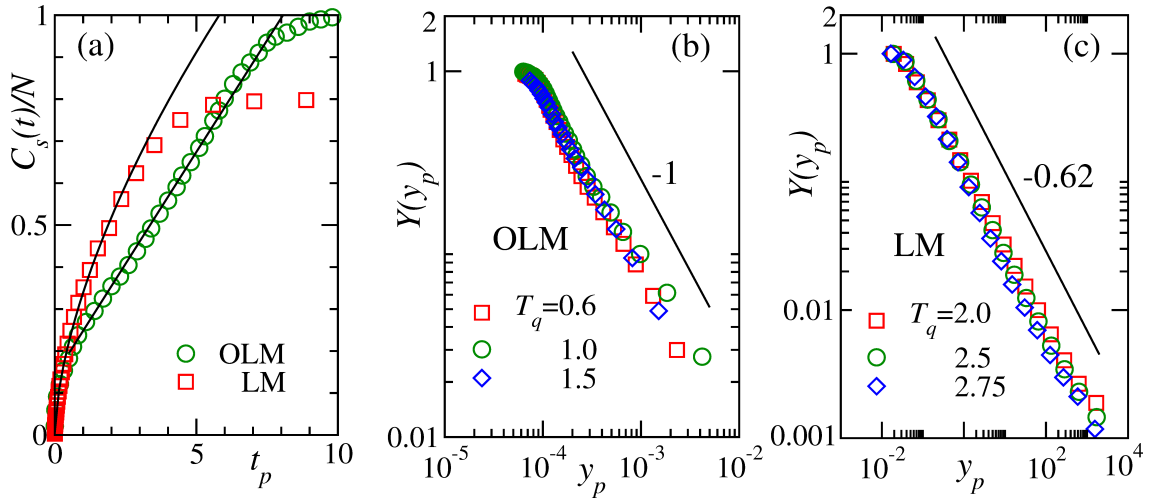}
\caption{\label{fig2} (a) Plots of the average cluster size $C_s(t)/N$ as function of time for the two models 
presented in Fig.\ \ref{fig1} (a). To make both the data visible on the same plot, we divide the time axis by 
a factor $m$ to obtain $t_p=t/m$, where $m=1\times10^{6}$ and $3.5\times 10^{6}$, respectively, for OLM and LM. 
The solid lines there are fits to the form \eqref{cl_growth} with $\alpha_c=0.98$ for OLM and $\alpha_c=0.62$ for LM. 
The plots in (b) and (c) demonstrate the scaling exercise, respectively, for OLM with $\alpha_c=1.0$ and LM with 
$\alpha_c=0.62$, showing that data for $C_s(t)$ at different quench temperatures $T_q$ 
can be collapsed onto a master curve using a nonuniversal metric factor in the scaling variable. 
The solid lines there represent the corresponding $Y(y_p) \sim y_p^{-\alpha_c}$ behavior.}
\end{figure}
\par
We now move on to a comparative study of the scaling of the cluster growth between the two models. References\ \cite{Suman_EPL,Suman_SM} 
provide details of the cluster identification method and subsequent calculation of $C_s(t)$ for OLM. On the other hand, 
for LM it is convenient to use $C_s(t) \equiv \ell(t)^3$, where $\ell(t)$ is obtained from the decay of the equal-time 
two-point density-density correlation function, for which we refer to Ref.\ \cite{Henrik}. However, here, for convenience, we use $C_s(t)$ as a 
notation for both the models. In Fig.\ \ref{fig2} (a), we show the time dependence of $C_s(t)$ for OLM and LM. In coarsening kinetics 
of binary mixtures \cite{Suman_rapid} such time dependence of the relevant length scale can be described correctly when one 
considers an off-set in the scaling ansatz. Similarly, it was later proved to be appropriate for the cluster growth during the collapse of 
a polymer \cite{Suman_EPL}. Following that one writes down the scaling ansatz as 
\begin{eqnarray}\label{cl_growth}
C_s(t)=C_0+At^{\alpha_{c}},
\end{eqnarray}
where $C_0$ corresponds to the cluster size after crossing over from the initial cluster formation stage, and 
$A$ is a temperature-dependent amplitude. The solid lines in Fig.\ \ref{fig2} (a) are fits to the form \eqref{cl_growth} 
yielding $\alpha_c =0.98(4)$ and $0.62(5)$, respectively, for OLM and LM. 
\par
As a step further, we verify the robustness of the growth by studying the dependence of cluster growth on the quench temperature $T_q$. 
For this we acquire data at different $T_q$ and perform scaling analyses based on 
nonequilibrium finite-size scaling (FSS) arguments \cite{Suman_rapid}. An account of the FSS formulation in the present context 
can be found in Ref.\ \cite{Suman_SM}. In brief, one introduces in the growth ansatz \eqref{cl_growth} a 
scaling function $Y(y_p)$ as 

\begin{eqnarray}\label{FS_func_cl}
C_s(t)-C_0=(C_{\max}-C_0)Y(y_p),~~ \textrm{i.e.,}~~ Y(y_p)=(C_s(t)-C_0)/(C_{\max}-C_0),
\end{eqnarray}
where $C_{\max} \sim N$ is the maximum cluster size a finite system can attain. In order to account for the 
temperature-dependent amplitude $A(T_q)$, one uses the scaling variable 

\begin{eqnarray}\label{FS_variable_T}
y_p= f_s(N-C_0)^{1/\alpha_{c}}/(t-t_0),~ ~\textrm{where} ~~f_s=\left[A(T_{q,0})/A(T_q)\right]^{1/\alpha_c}.
\end{eqnarray}
The metric factor $f_s$ is introduced for adjusting the nonuniversal amplitudes $A(T_q)$ at different $T_q$ \cite{Suman_SM}. Here, in addition to $C_0$ one also 
uses the crossover time $t_0$ from the initial cluster formation stage. 
A discussion of the estimation of $C_0$ and $t_0$ can be found in Refs.\ \cite{Suman_SM,Henrik}. While performing the exercise 
we tune the parameters $\alpha_c$ and $f_s$ to obtain a data collapse along with the $Y(y_p) \sim y_p^{-\alpha_c}$ behavior 
in the finite-size unaffected region. In Fig.\ \ref{fig2} (b) and (c), we demonstrate such scaling exercises 
for OLM and LM with $\alpha_c=1.0$ and $0.62$, respectively. For $f_s$, we use the reference temperature $T_{q,0}=1.0$ and $2.0$, respectively, 
for OLM and LM. The collapse of data for different $T_q$ and consistency with the corresponding $y^{-\alpha_c}$ behavior in both plots 
suggest that the growth is indeed quite robust and can be described by a single finite-size scaling function with nonuniversal metric factor $f_s$ 
in the scaling variable. However, $\alpha_c$ in OLM is larger than for LM, a fact in concurrence with the values 
of $z$ estimated previously, and thus supporting the argument that $z \sim 1/\alpha_c$. 
\begin{figure}[t!]
\centering
\includegraphics*[width=0.90\textwidth]{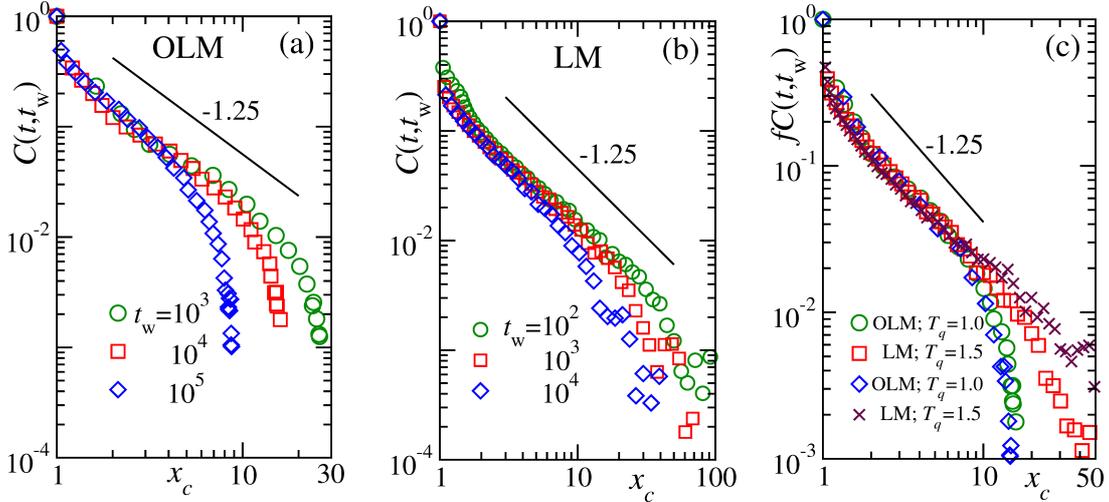}
\caption{\label{fig3} (a) Plot showing the scaling of $C(t,t_w)$ as function of $x_c=C_s(t)/C_s(t_w)$ in OLM, for three 
different waiting times $t_w$, at $T_q=1$. The solid line shows the behavior \eqref{FH-bound} with $\lambda_C=1.25$. (b) Same as (a) but 
for LM at $T_q=1.5$. (c) Plot showing that aging scaling at different $T_q$ for the two models can be described 
by a single master-curve behavior. The solid line here also corresponds to Eq.\ \eqref{FH-bound} with $\lambda_C=1.25$. 
Note that $C(t,t_w)$ is multiplied by a factor $f$ to make them collapse onto the same curve.
For OLM $t_w=10^4$ whereas for LM $t_w=10^3$.}
\end{figure}
\par
Next we compare the results from aging and related scaling during the collapse of the two model polymers. For OLM 
it is advised \cite{Suman_aging} to construct an autocorrelation function by assigning $O_i=\pm1$ in \eqref{auto_cor}, based on whether 
a monomer is inside a cluster or not, an analogy of the local density criterion. On the other hand, for LM we can define $O_i=\pm1$ 
directly from the local density. For a description of the construction of such autocorrelation function 
we refer to Refs.\ \cite{Suman_aging,Suman_SM} and Ref.\ \cite{Henrik}, respectively, for OLM and LM. 
In Fig.\ \ref{fig3} (a) and (b), we demonstrate the simple aging behavior by showing the scaling 
of $C(t,t_w)$ for three different $t_w$, with respect to the scaling variable $x_c$. Data for both models 
show consistency with Eq.\ \eqref{FH-bound}, having an exponent $\lambda_C=1.25$. The value indicates that $\lambda_C$ 
indeed follows the bound \eqref{poly-bound}, which on transforming to numerical values \cite{Suman_aging} 
provides $0.762\,791\le \lambda_C \le 1.525\,582$. 
For LM it has been shown \cite{Henrik} that scaling with respect to $t/t_w$ may 
lead to the misconclusion of the presence of sub-aging, which merely is the result of fitting with a complex variable, as 
pointed out previously in ordering of diluted ferromagnets \cite{Pleimling}. Lastly, in Fig.\ \ref{fig3} (c), 
we show that the data for the two models can be collapsed onto a single master-curve behavior, irrespective of $T_q$. 
The multiplier $f$ on the $y$-axis, \cite{Suman_JPCM,Suman_SM} 
adjusts different amplitudes, $A_C$, for different $T_q$ as well as models. Thus, unlike the growth exponent, 
the nonequilibrium autocorrelation exponent $\lambda_C$ is rather universal.

\section{Conclusion}
We have compared results from kinetics of the collapse transition in an off-lattice and a lattice homopolymer model. 
The smaller value of the exponent $z$, governing the scaling of collapse time, in the off-lattice model than 
in the lattice model suggests that the dynamics is faster in the former. This, perhaps, is controlled by the exponent 
$\alpha_c$ ($\alpha_c \sim 1/z$), characterizing the cluster-growth stage which seems to 
be the rate limiting stage of the overall collapse process. While the off-lattice 
model yields a linear growth ($\alpha_c \approx 1$), in the lattice model the growth is slower ($\alpha_c\approx 0.62$), which could 
be attributed to the topological constraints one experiences in a lattice model. On the other hand, surprisingly, 
both the models show evidence of simple aging scaling having the same autocorrelation exponent $\lambda_C \approx 1.25$, 
thus implying that the aging scaling is rather universal. This allowed us to demonstrate that scaling of the 
autocorrelation functions for the two models can be described by a single master curve. 
\ack
The work was funded by the Deutsche Forschungsgemeinschaft (DFG) under Grant Nos.\ JA 483/33-1 and SFB/TRR 102 (project B04), 
and further supported by the Leipzig Graduate School of Natural Sciences ``BuildMoNa,''
the Deutsch-Franz\"osische Hochschule (DFH-UFA) through the Doctoral College ``${\mathbb L}^4$'' under Grant No.\ CDFA-02-07, 
and the EU Marie Curie IRSES network DIONICOS under Contract No.\ PIRSES-GA-2013-612707.

\end{document}